\documentclass[times,sort&compress,3p]{elsarticle}
\journal{Journal of Multivariate Analysis}
\usepackage{amsmath, amssymb, enumerate, amsthm}

\usepackage{natbib}
\usepackage{epsfig}
\usepackage{subfigure}
\usepackage{amsfonts}
\usepackage{xcolor}
\usepackage{bm}
\usepackage{array}
\usepackage{caption}
\usepackage{hyperref}

\hypersetup{colorlinks=true, linkcolor=black, citecolor=black, urlcolor=black}

\newtheorem{theorem}{Theorem}

\newtheorem{corollary}{Corollary}
\newtheorem{proposition}{Proposition}
\newtheorem{remark}{Remark}

\newcommand{\sig}{\sigma^2}
\newcommand{\hr}{h_{\rho}}

\newcommand{\maxmc}{\max_{\{j:j\in\calm^\complement\}}}
\newcommand{\minmc}{\min_{\{j:j\in\calm^\complement\}}}
\newcommand{\tcorr}{\rm{corr}}
\newcommand{\hcorr}{\widehat{\tcorr}}

\newcommand{\rj}{\mathbf{r}_j}

\newcommand{\xm}{X_{\calm}}
\newcommand{\simd}{\,{\buildrel \text{d} \over \sim\,}}

\newcommand{\limd}{\rightsquigarrow}

\newcommand{\RNum}[1]{\uppercase\expandafter{\romannumeral #1\relax}}
\newcommand{\wc}{{}\cdot{}}

\newcommand{\PP}[1]{\Pr\left(#1\right)}

\newcommand{\npn}{(p-s)^{-2/(n-s-2)}}
\newcommand{\apn}{a(p, n, s)}
\newcommand{\bpn}{b(p,n,s)}
\newcommand{\cpn}{c(p, n, s)}

\newcommand{\xcrit}{x_{\alpha}(p, n, s)}
\newcommand{\tpn}{T_{\calm}}
\newcommand{\rpn}{R_{\calm}}

\newcommand{\tapn}{\widetilde{a}(p,n,s)}
\newcommand{\tbpn}{\widetilde{b}(p,n,s)}

\newcommand{\Betans}{\calb[1/2, (n-s-2)/2]}
\newcommand{\hbeta}{\hat{\beta}}
\newcommand{\veps}{\varepsilon}

\newcommand{\infp}{p \rightarrow \infty}

\newcommand{\cala}{\mathcal{A}}
\newcommand{\calm}{\mathcal{M}}
\newcommand{\caln}{\mathcal{N}}

\newcommand{\calb}{\mathcal{B}}

\newcommand{\bX}{\mathbf{X}}
\newcommand{\bx}{\mathbf{x}}
\newcommand{\by}{\mathbf{y}}
\newcommand{\bz}{\mathbf{z}}
\newcommand{\br}{\mathbf{r}}

\newcommand{\bveps}{\boldsymbol{\veps}}
\newcommand{\bbeta}{\boldsymbol{\beta}}

\newcommand{\tbr}{\tilde{\br}}

\newcommand{\onen}{\mathbf{1}_n}
\newcommand{\projm}{P_{\calm}}
\newcommand{\proja}{P_{\cala_{k-1}}}
\newcommand{\projone}{\onen\onen^\top/n}
\newcommand{\barxj}{\bar{x}_j\mathbf{1}_n}

\begin{document}

\begin{frontmatter}		

\title{Efficient test-based variable selection for high-dimensional linear models}

\author[1]{Siliang Gong\corref{cor}}
\cortext[cor]{Corresponding author}
\ead{siliang@live.unc.edu}

\author[1]{Kai Zhang}
\ead{zhangk@email.unc.edu}

\author[1,3]{Yufeng Liu}
\ead{yfliu@email.unc.edu}

\address[1]{Department of Statistics and Operations Research, University of North Carolina at Chapel Hill, Chapel Hill, NC 27599}
\address[3]{Depts of Genetics and Biostatistics, Carolina Center for Genome Science, University of North Carolina at Chapel Hill, Chapel Hill, NC 27599}

	\begin{abstract}
		Variable selection plays a fundamental role in high-dimensional data analysis. Various methods have been developed for variable selection in recent years. Well-known examples are forward stepwise regression (FSR) and least angle regression (LARS), among others. These methods typically add variables into the model one by one. For such selection procedures, it is crucial to find a stopping criterion that controls model complexity. One of the most commonly used techniques to this end is cross-validation (CV) which, in spite of its popularity, has two major drawbacks: expensive computational cost and lack of statistical interpretation. To overcome these drawbacks, we introduce a flexible and efficient test-based variable selection approach that can be incorporated into any sequential selection procedure. The test, which is on the overall signal in the remaining inactive variables, is based on the maximal absolute partial correlation between the inactive variables and the response given active variables. We develop the asymptotic null distribution of the proposed test statistic as the dimension tends to infinity uniformly in the sample size. We also show that the test is consistent. With this test, at each step of the selection, a new variable is included if and only if the $p$-value is below some pre-defined level. Numerical studies show that the proposed method delivers very competitive performance in terms of variable selection accuracy and computational complexity compared to CV.
	\end{abstract}
	
	\begin{keyword}
Cross-validation \sep High-dimensional testing \sep Maximal absolute correlation \sep Variable selection.	
	\end{keyword}
	
\end{frontmatter}

\section{Introduction}\label{sec:intro}

Thanks to technological advancement, high-dimensional data are now prevalent in science. Unfortunately, traditional techniques such as ordinary least squares cannot be applied directly to these high-dimensional settings, where the number of variables is typically much larger than the sample size. Furthermore, it is often the case that only a few candidate predictors are truly relevant to the response \citep{fan14big}. In other words, the inherent high-dimensional model is sparse. It is then crucial to identify such variables, whence the important problem of variable selection arises.

In the context of linear regression, various variable selection procedures have been intensively investigated in the past decades. One example is forward stepwise regression (FSR); see \cite{hocking76} for a review. Another well-known example is the least absolute shrinkage and selection operator (LASSO) proposed by Tibshirani~\cite{tib96}. The LASSO is a sparse regularized least squares method for linear regression, which imposes the $L_1$ penalty on regression coefficients. Efron et al.~\cite{efron04lars} proposed the least angle regression (LARS) method, which can compute efficiently the entire solution path of the LASSO with respect to the tuning parameter. As shown in~\cite{efron04lars}, LARS is also less greedy than FSR, and the solution paths of LARS and LASSO are piecewise linear. Many other sparse penalized methods have been proposed in the literature, e.g., the Dantzig selector~\citep{candes07} and the smoothly clipped absolute deviation (SCAD) penalty~\citep{fan01noncon}.

The variable selection methods discussed above usually involve a penalty parameter which controls the complexity of the resulting model. In practice, cross-validation (CV) is a commonly used technique for selecting the penalty parameter. However, CV is computationally inefficient. Moreover, it is based on minimizing in-sample prediction errors, and thus does not have a clear inferential meaning.
Besides CV, another class of model selection approaches is based on hypothesis testing. For example, \cite{goeman06} and \cite{zhong11} focused on testing the regression coefficients globally.

Other testing schemes have been implemented adaptively in sequential selection procedures. For example, Lockhart et al. \cite{lockhart14sig} proposed the covariance test statistic for the LASSO. Another example is the truncated Gaussian (TG) test~\citep{taylor14exact} developed for LARS, FSR and LASSO. While these methods are specifically designed for particular variable selection procedures, Fithian et al.~\cite{fithian15} introduced a general framework for testing the goodness of fit that applies to FSR, LARS and LASSO. However, their tests are developed separately for FSR and LARS (LASSO). In addition, the method of Fithian et al.~\cite{fithian15} requires MCMC sampling for the null distribution, which can be time consuming.

For LARS, FSR and LASSO, test-based approaches are applicable because these procedures are sequential in nature: typically, only one variable is added into the model at each step (though the LASSO can sometimes include steps in which variables are dropped). Therefore, tests can be conducted at each step of the selection procedure. One can further develop some stopping criterion based on the $p$-values associated with these hypothesis tests.

Another common feature of these procedures is that at each step, a variable is selected if, among all inactive variables, it has the largest absolute sample correlation with the current residuals, i.e., the difference between the response and its estimates from the previous step. However, such a large sample correlation can be spurious. Indeed in situations where the number of predictors is large compared to the sample size, it may happen that the response is theoretically independent from all of them and yet some of these predictors appear to be highly correlated with the response simply by chance. This phenomenon can be particularly severe in high-dimensional problems. As mentioned, e.g., by Fan et al.~\citep{fan14big}, the maximal correlation observed in a sample of fixed size $n$ between a response and independent covariates can be close to 1 if the number $p$ of such covariates is sufficiently large.

In this paper, we introduce an efficient high-dimensional test-based variable selection method. We focus on the variable selection problem under the sparse linear model setting. Motivated by the spurious correlation issue discussed above, we construct a test statistic based on the maximal absolute sample partial correlation between the inactive covariates and the response conditioning on the active covariates at each step of the procedure. Our null hypothesis assumes that the remaining variables are conditionally independent of the response given the active variables. Based on the null distribution of the test statistic, we can detect whether there exist important covariates for the response in the inactive set. We further develop a stopping criterion from the $p$-values.

There are three key advantages to the proposed method, namely:
\begin{itemize}
\item [(i)]
The method is flexible: the proposed tests and stopping criterion can be incorporated into any sequential selection procedure, such as the aforementioned LARS, LASSO and FSR.
\item [(ii)]
The method is much more computationally efficient than CV, especially when $p$ is large.
\item [(iii)]
The method can accommodate arbitrarily large $p$, since the asymptotic null distribution of the test statistic is developed as $p \to \infty$ uniformly in $n$.
\end{itemize}

This paper is organized as follows. In Section~\ref{sec:gtest}, we formulate the null hypothesis and introduce the corresponding test statistic for the proposed method. In Sections~\ref{sec:iid} and \ref{sec:power}, we discuss the asymptotic null distribution and power of our test statistic with independent covariates, respectively, and we extend the results for equally correlated covariates in Section~\ref{sec:cor}. In Section~\ref{sec:pt}, we introduce the permutation test for covariates with arbitrary correlation structure. In Section~\ref{sec:seqtest}, we incorporate our hypothesis testing approach into sequential variable selection procedures. In Section~\ref{sec:ns}, we demonstrate the performance of the new method through three simulation studies and a microarray data study. Proofs and additional simulation results are given in the Online Supplement.

\section{Global test to control spurious correlation}\label{sec:md}

\subsection{Global null for testing significant variables}\label{sec:gtest}

Consider the linear model
\begin{equation}\label{eq:lm}
Y = \bX^\top\bbeta+\veps,
\end{equation}
where $Y$ is the response variable, $\bX = (X_1,  \ldots, X_p)^\top$ is a $p$-dimensional covariate vector, $\bbeta= (\beta_1,  \ldots, \beta_p)^\top$ is the unknown coefficient vector which may be sparse, and $\veps$ is a random noise from $\caln(0, \sig)$ with $\sig$ unknown. For now we assume that $\bX$ is from a $p$-dimensional Gaussian distribution with some unknown covariance matrix $\Sigma$. We will discuss the non-Gaussian case in the numerical studies. Let $\by = (y_1, \ldots, y_n)^\top$ and $\bx_j = (x_{1j}, \ldots, x_{nj})^\top$ respectively stand for the vectors of independent observations from $Y$ and $X_j$, with $j \in \{ 1, \ldots , p\}$.

For variable selection problems, the primary goal is to recover the support set of $\bbeta$, which is the index set of non-zero components of the coefficient, denoted $\calm^*$. Suppose we are given a candidate set $\calm$, which includes the indices of all selected variables, and that we want to know whether there are remaining important covariates in $\calm^\complement$. We then need to test
\begin{equation}\label{eq:orignull}
\mathcal{H}_0: \calm^*\subseteq \calm.
\end{equation}

The following proposition demonstrates that under \eqref{eq:lm} and the Gaussian assumption, we can convert the above hypothesis into the problem of testing the conditional independence between $Y$ and the $X_j$s with $j\in\calm^\complement$.

\begin{proposition}\label{prop:nulleq}
	Suppose that $\bX = (X_1,  \ldots, X_p)^\top$ has a multivariate Gaussian distribution and the response $Y$ is generated from the linear model \eqref{eq:lm}. If $\calm$ is a subset of $\{1,  \ldots, p\}$, then $\calm^*\subseteq \calm$ if and only if $Y$ is independent of all $X_j$s for $j\in\calm^\complement$ conditional on $X_\calm$.
\end{proposition}

Proposition \ref{prop:nulleq} guarantees that testing \eqref{eq:orignull} is equivalent to the following null hypothesis:
\begin{align}\label{eq:null}
\mathcal{H}_0^{\calm}: \text{Given }X_\calm, Y \text{ is independent of all } X_j\text{s for } j \in \calm^\complement.
\end{align}

Unless the noise is very strong, the correlation between an important covariate and the response should be stronger than the maximal spurious correlation. In fact, many existing variable selection methods, such as the LASSO and FSR, select variables that maximize the absolute marginal correlation between the covariates and the response or the current residuals. Moreover, it is easy and efficient to obtain the maximal absolute correlation, even if the dimension~$p$ is high. Therefore, studying the distribution of the maximal absolute correlation under the null hypothesis~\eqref{eq:null} can help discover true important covariates among the candidate predictor variables.

We cannot directly test \eqref{eq:null} based on the correlation between $Y$ and $X_j$ because they can be both correlated with $X_i$ for some $i\in \calm$. In classical regression, the partial correlation is commonly used to test conditional independence given a controlling variable. Motivated by that observation, we develop our test statistic based on the sample partial correlation between $\{X_j : j\in \calm^\complement\}$ and $Y$ conditioning on $X_\calm$. We first regress $\{X_j: j\in \calm^\complement\}$ and $Y$ onto $X_{\calm}$, respectively; we then obtain the regression residual vectors
\begin{equation}
\label{eq:res}
\br_j = (I-P_{\calm})\bx_j,\quad j\in\calm^\complement, \qquad \br = (I-P_{\calm})\by,
\end{equation}
where $P_{\calm}=\xm(\xm^\top\xm)^{\dagger}\xm^\top$ is the projection onto the column space of $\xm$. Here $\xm$ consists of the columns of $X$ indexed by $\calm$ and a vector column of $1$s, so that all residual vectors have zero mean, and $A^{\dagger}$ denotes the Moore--Penrose pseudo-inverse of a matrix $A$. We then compute the maximal absolute sample correlation between $\{\br_j:  j\in\calm^\complement\}$ and $\br$. In this way, we define our test statistic as
\begin{equation} \label{eq:maxcorr}
\rpn = \maxmc |\hcorr(\br_j, \br)|,
\end{equation}
where $\hcorr(\br_j, \br)$ is the Pearson sample correlation between $\br_j$ and $\br$. Note that the distribution of $\rpn$ depends on $n, p$ and $s$, but for simplicity we omit them in the notation for $\rpn$. Since both $\br_j$ and $\br$ have zero mean, we can write
\begin{equation*}
\rpn = \maxmc {\frac{|\langle \br_j, \br \rangle|}{\|\br_j\| \|\br\|} } ,
\end{equation*}
where $\langle \wc, \wc\rangle$ is the inner product of two vectors and $\|\cdot\|$ represents the $L_2$ norm. Moreover, note that our test statistic does not depend on the mean and variance of the covariates or the response.

To gain insight into the proposed test statistic, we start from a special case where $\calm=\emptyset$. The properties of the Pearson sample correlation have been intensively studied under the classical setting $n>p$. In particular, it has been shown that when $X_j$ and $Y$ are independent Gaussian random variables, $|\hcorr(X_j, Y)|^2\sim \calb[1/2, (n-2)/2]$; see, e.g.,~\citep{muirhead82}. Therefore, the magnitude of each $\hcorr(X_j, Y)$ cannot be too large. However, by taking maxima, $\rpn$ will be larger as $p$ increases. In fact, for a fixed sample size $n$, under \eqref{eq:null}, $\rpn$ can get close to $1$ as $p \to \infty$; see, e.g.,~\citep{fan14big}. The phenomenon of irrelevant covariates being highly correlated with the response is referred to as ``spurious correlation'', which challenges variable selection and may lead to false scientific discoveries. Thus it is important to study the distribution of $\rpn$, especially for high-dimensional problems.

In what follows, we discuss the asymptotic null distribution (Section \ref{sec:iid}) and power (Section \ref{sec:power}) of $\rpn$ respectively for the situation where the $X_j$s are independent random variables. We discuss the situation where the covariates are dependent in Section \ref{sec:cor}.

\subsection{Null distribution of the test statistic with independent covariates}\label{sec:iid}

The limiting distribution of the maximal absolute sample correlation has been investigated recently under the assumption of independent Gaussian covariates; see Theorem~II.4 in \citep{zhang15pack}. The latter paper focuses on the global null hypothesis that $Y$ is independent of the $X_j$s, which is a special case of \eqref{eq:null} with $\calm = \emptyset$. We expand the results to a more general setting and derive the exact asymptotic distribution of the proposed test statistic under \eqref{eq:null}, as described in the following theorem

\begin{theorem}
	\label{thm:corr}
	Suppose we observe a random sample of size $n$ from the linear model \eqref{eq:lm} and we further assume that the $X_j$s are independent. Let $\calm$ be a candidate set with cardinality $|\calm| = s< n-2$ and $\rpn$ be defined as in~\eqref{eq:maxcorr}. Define
	$$
	\apn=1-\npn\cpn, \quad \bpn=\frac{2}{n-s-2}\, \npn\cpn,
	$$
	where $\cpn = \{2^{-1}(n-s-2)\Betans\sqrt{1-\npn}\}^{2/(n-s-2)}$
	is a correction factor with $\calb(s,t)$ being the Beta function. Then under the null hypothesis \eqref{eq:null}, for all $x \in \mathbb{R}$,
	\begin{equation*}\label{eq:mcor}
	\lim_{p \to \infty} \sup_{n \ge s+3} \left|\Pr \left\{ \frac{\rpn^2 - \apn}{\bpn} < x\right\} -F_{n,s}(x)\right| = 0,
	\end{equation*}
	where
	\begin{equation}\label{eq:asymp}
	F_{n,s}(x) = \exp\left\{ -\left(1-\frac{2}{n-s-2} x\right)^{{(n-s-2)}/{2}}\right\} \mathbf{1} \left(x \le \frac{n-s-2}{2}\right)+ \mathbf{1} \left(x>{\frac{n-s-2}{2}}\right).
	\end{equation}	
\end{theorem}

\begin{remark}
\em
The convergence in Theorem~\ref{thm:corr} is with respect to $p$ instead of $n$, making it possible to test models where $p\gg n$. Therefore, the proposed test statistic is applicable to high-dimensional or ultra-high-dimensional problems. In addition, the convergence is uniform for any $n\ge s+3$, and thus ensures finite-sample performance.		
\end{remark}

With the results in Theorem \ref{thm:corr}, we can further compute the $p$-value associated with the null hypothesis $\eqref{eq:null}$. Let $r_{\calm}$ denote the observed value of $\rpn$. Then the $p$-value of $\rpn$ for \eqref{eq:null} is
\begin{equation}\label{eq:piid}
p(r_{\calm}) =1 - F_{n,s}\left\{ \frac{r_{\calm}-\apn}{\bpn}\right\},
\end{equation}
with $F_{n,s}$ as specified in Theorem~\ref{thm:corr}. If the $p$-value is small, it is likely that at least one variable from~$\{X_j: j\in \calm^\complement \}$ is correlated with the response. Therefore we can construct a stopping criterion based on $p$-values in sequential selection procedures. We will provide a detailed discussion in Section~\ref{sec:seqtest}.

Our test statistic can be connected to the conventional $t$-test for testing whether the population correlation is zero. The $t$-statistic is defined as $t = r\sqrt{(n-2)/(1-r^2)}$, where $r$ is the Pearson sample correlation between two Gaussian random variables. Motivated by that connection, we also develop a maximal $t$-statistic corresponding to the proposed test statistic $\rpn$. The maximal $t$-statistic is
\begin{equation}\label{eq:tstat}
\tpn = \sqrt{\frac{(n-s-2)\rpn^2}{1-\rpn^2}}.
\end{equation}

Analogous to the results in Theorem~\ref{thm:corr}, we derive next the asymptotic null distribution of $\tpn$.

\begin{corollary} \label{col:tstat}
Consider the same setting as in Theorem \ref{thm:corr}, and let $\tpn$ be defined as in \eqref{eq:tstat}. Then, for all $x \in \mathbb{R}$, uniformly for any $n\ge s+3$,
\begin{equation*}
\lim_{p \to \infty} \Pr\left\{ \frac{T_{\calm}-\tapn}{\tbpn}<x\right\} = F_{n,s}(x),
\end{equation*}
where $\tapn = \sqrt{\{(n-s-2)\apn\}/\{1-\apn\}}$, $\tbpn = [(n-s-2)\apn\{1-\apn \}]^{-1/2}$ with $\apn$ given in Theorem~\ref{thm:corr}, and $F_{n,s}(x)$ as in~\eqref{eq:asymp}.
\end{corollary}

Our simulation results show that the difference between $p$-values obtained from $\rpn$ and $\tpn$ is negligible. Moreover, when the covariates are correlated, the null distribution of $\rpn$ is easier to approximate, which will be discussed in Section~\ref{sec:cor}. Therefore we develop our test-based procedure with $\rpn$ instead of~$\tpn$.

\subsection{Asymptotic power with independent covariates}\label{sec:power}

In this section, we still focus on independent Gaussian covariates. We analyze the asymptotic power of $\rpn$ by considering the following alternative hypothesis:
\begin{equation} \label{eq:alt}
\mathcal{H}_1: \text{Conditionally on }X_{\calm}, \text{ there exists at least one  } j\in \calm^\complement \text{ such that } Y \text{ is correlated with } X_j.
\end{equation}

In the following theorem we show that under \eqref{eq:alt}, the asymptotic power of the proposed test statistic $\rpn$ is 1.

\begin{theorem}\label{thm:power}
	Suppose we have the linear model \eqref{eq:lm} and assume that the $X_j$s are independent Gaussian variables. Then under the alternative hypothesis \eqref{eq:alt}, as ${\ln p}/n\rightarrow 0$ and $n\rightarrow \infty$, $\Pr \{\rpn \ge \xcrit|\mathcal{H}_1\} \longrightarrow 1$, where $\xcrit$ is the critical value of $\mathcal{H}_0^{\calm}$ at significance level $\alpha$.
\end{theorem}

Theorem~\ref{thm:power} shows the consistency of our dependency test based on the proposed test statistic when at least one covariate is correlated with the response under the linear model setting.

\subsection{Null distribution of the test statistic with equally correlated covariates}\label{sec:cor}

In Theorem~\ref{thm:corr} we have derived the exact asymptotic distribution of $\rpn$ under \eqref{eq:null} when the covariates are independent Gaussian variables. When the $X_j$s have an arbitrary correlation structure, it is difficult to obtain similar results. We can point to some results in classical extreme-value theory; see, e.g., Chapter~3.8 in~\citep{galambos78}. In particular, if $U_1, \ldots, U_n$ is a stationary Gaussian sequence with zero expectation and unit variance, then the limiting distribution of $W_n = \max(U_1, \ldots, U_n)$ only depends on the limiting behavior of ${r_m}/{\ln(m)}$, where $r_m = {\rm E} (U_{i}U_{i+m})$ is the correlation between $U_i$ and $U_{i+m}$. Note that due to the stationarity assumption, $r_m$ does not change with respect to $i$. More specifically, if there is another zero-mean, unit-variance stationary Gaussian sequence $U_1', \ldots, U_n'$ that has equal pairwise correlation $r = r(n)$, and ${r(n)}/{\ln(n)}$ has the same limiting form as ${r_m}/{\ln(m)}$, then $W^{'}_n = \max(U_1', \ldots, U_n')$ has the same asymptotic distribution as $W_n$ when $n \to \infty$. Inspired by that result, we focus on analyzing the null distribution of $\rpn$ when $X_1, \ldots, X_p$ are equally correlated, i.e., $\tcorr(X_i, X_j)=\rho$ with $-1/(p-1)\le\rho\le 1$ for all $i \ne j$.

Without loss of generality, we assume that each of the $X_j$s has zero mean and unit variance. Under the equal correlation assumption, it is well known that we can decompose $X_j$ into a linear combination of iid standard Gaussian random variables $Z_1, \ldots, Z_p$, i.e.,
\begin{equation}\label{eq:decomp}
X_j = \sqrt{1-\rho}\, Z_j + \hr\frac{1}{\sqrt{p}}  \sum_{i=1}^p Z_i,
\end{equation}
where $\hr = \{\sqrt{1+(p-1) \rho}-\sqrt{1-\rho}\}/{\sqrt{p}}$.
In fact, we can also replace $p$ by $p-s$ in \eqref{eq:decomp} such that each of $\{X_j: j\in \calm^\complement\}$ is decomposed into a linear combination of $p-s$ iid Gaussian random variables. However, under  high-dimensional sparse model settings, $p\gg s$. Hence the two decompositions are almost the same. For computational simplicity, we consider using $p$ instead of $p-s$.

Let $\bz_j=(z_{j1}, \ldots, z_{jn})^\top$ be $n$ independent samples of $Z_j$ and $\tbr_j = (I-\projm)\bz_j$ be the residuals from projecting $\bz_j$ onto the column space of $\xm$. It follows from \eqref{eq:decomp} that
\begin{equation*}
\br_j = \sqrt{1-\rho} \, \tbr_j +  \hr\frac{1}{\sqrt{p}}  \sum_{i=1}^p \tbr_i.
\end{equation*}
Hence we have
\begin{equation*}
\langle\br_j, \br\rangle = \sqrt{1-\rho} \, \langle\tbr_j, \br\rangle + \hr \left\langle p^{-1/2}  \sum_{i=1}^p \tbr_i, \right\rangle,
\end{equation*}
where $\br_j$ and $\br$ are defined as in \eqref{eq:res}.

Recall that by assumption, $\text{var}(Z_j) = \text{var}(X_j)=1$. Thus conditioning on $\xm$, we have
$$
\|\br_j\|^2\simd \chi^2_{n-s-1},\quad \|\tbr_j\|^2\simd \chi^2_{n-s-1}, \quad \|p^{-1/2}\sum_{i=1}^p \tbr_i\|^2\simd \chi^2_{n-s-1}.
$$
For moderately large $n$, we can approximate $\hcorr(\br_j, \br) = \langle \br_j, \br\rangle/ (\|\br_j\|\|\br\|)$ by
\begin{equation*}
\hcorr (\br_j, \br) \approx \sqrt{1-\rho} \, \hcorr(\tbr_j, \br)+\hr \hcorr \left(p^{-1/2}\sum_{i=1}^p \tbr_i, \right).
\end{equation*}
Taking the maximum on both sides, we find
\begin{equation}\label{eq:cordec}
\maxmc \hcorr(\br_j, \br) \approx  \sqrt{1-\rho}\maxmc \hcorr(\tbr_j, \br) + \hr \hcorr \left(p^{-1/2}\sum_{i=1}^p \tbr_i, \right).
\end{equation}

Under the null hypothesis \eqref{eq:null}, note that $\br = (I-\projm)\by = (I-\projm)\bveps$ and thus $\tbr_j = (I-\projm)\bz_j$ is conditionally independent of $\br$ given $\xm$ for all $j \in \{ 1, \ldots, p\}$. Hence the variables $\{|\hcorr(\tbr_j, \br)|^2:j\in\calm^\complement\}$ are independently distributed as $\Betans$ conditioning on $\xm$. Furthermore, from a property of the normal distribution,
$$
p^{-1/2} \sum_{i=1}^p Z_i \simd \caln(0,1).
$$
Thus the conditional distribution of  $|\hcorr(p^{-1/2}\sum_{i=1}^p \tbr_i, \br)|^2$ given $\xm$ is also $\Betans$. Therefore, the two terms on the right-hand side of \eqref{eq:cordec} have corresponding exact distributions. Letting $f_1, f_2$ be the densities of $\maxmc\hcorr(\tbr_j, \br)$ and $\hcorr(p^{-1/2}\sum_{i=1}^p \tbr_i, \br)$, respectively, we have
\begin{equation*}
f_1(x;p,n,s) =  p|x|f_B(x^2;n,s)\left\{\frac{1+\text{sign(x)}F_B(x^2;n,s)}{2}\right\}^{p-s-1},\quad
f_2(x;n,s) = |x|f_B(x^2;n,s),
\end{equation*}
where $f_B(x;n,s)$ and $F_B(x;n,s)$ are the density and the cumulative distribution function of $\Betans$, respectively.

It is known that when $\infp$, $\max (Z_1, \ldots, Z_p)$ and $Z_1 + \cdots + Z_p$ are independent; see, e.g., \citep{james07}. With asymptotic independence, the density $f_3(x;p,n,s)$ of $\maxmc\hcorr(\tbr_j, \br)$ can be approximated, for all $z \in [0,1]$, by
\begin{equation}\label{eq:conv}
f_3(z;p,n,s) \approx \int_{-\infty}^{\infty}\tilde{f}_1(z-x)\tilde{f}_2(x) dx,
\end{equation}
with $\tilde{f}_1(x) = \rho^{-1/2}f_1(\rho^{-1/2} x;p,n,s)$ and $\tilde{f}_2(x) = f_2(x/\hr;n,s)/\hr$. In practice, $\rho$ can be estimated by the average of pairwise correlations among the covariates. Let
$$
U_{\calm} = \maxmc\hcorr(\br_j, \br), \quad V_{\calm} = -\minmc\hcorr(\br_j, \br).
$$
Note that $\rpn = \max(U_{\calm}, V_{\calm} )$, where $U_{\calm}$ and $V_{\calm}$ have identical distributions, but are not independent.

Due to the dependence between $U_{\calm}$ and $V_{\calm}$, it is difficult to derive the distribution of $\rpn$ and the corresponding $p$-value when we use $\rpn$ as the test statistic. One possible way to tackle this problem is to take $U_{\calm}$ or $V_{\calm}$ as the test statistic instead. However, the resulting test might not be powerful enough. For example, when the true model is $Y = -X_1+\veps$, it is difficult to reject the null hypothesis~\eqref{eq:null} based on the null distribution of $U_{\calm}$. Similarly, if the true model is $Y = X_1+\veps$, then using $V_{\calm}$ as the test statistic might be unable to detect $X_1$. However, note that if the null hypothesis does not hold, i.e., there are important variables remaining in $\calm^\complement$, it can be expected that the tail probability of $\rpn$ will be very small. It can then be approximated by
\begin{equation}\label{eq:paprox}
\PP{\rpn \ge x} \approx \PP{U_{\calm} \ge x}+\PP{V_{\calm} \ge x}
 = 2\PP{U_{\calm} \ge x}.
\end{equation}

Since $\PP{\rpn \ge x}\in [\PP{U_{\calm} \ge x}, 2\PP{U_{\calm} \ge x}]$ always holds, if $2\PP{U_{\calm} \ge x}$ is small, $\PP{\rpn \ge x}$ will also be very small, which implies that the null hypothesis may be rejected. Therefore we can compare $2\PP{U_{\calm} \ge x}$ with a pre-specified constant $c$ to determine which test statistic, $\rpn$ or $U_{\calm}$ to use. In general, we propose to compute the $p$-value corresponding to \eqref{eq:null} in the following way:
\begin{equation}\label{eq:pcor}
p = \left\{
\begin{array}{ll}
\PP{\rpn \ge x}\approx 2\PP{U_{\calm}\ge x} & \text{if } 2\PP{U_{\calm}\ge x}\le c, \\
\PP{U_{\calm}  \ge x_1} & 	\text{otherwise,}\\
\end{array}
\right.
\end{equation}
where $x$ and $x_1$ represent the observed value of $\rpn$ and $U_{\calm}$, respectively, and
$$
\PP{U_{\calm} \ge t} \approx \int_{t}^{\infty} f_3(z;p,n,s)dz.
$$
The constant $c$ is essentially a parameter balancing the accuracy and conservatism of the resulting $p$-value. Specifically, if $c$ is too small, the $p$-value is then computed from $U_{\calm}$, which can be too conservative; if $c$ is too large, the  approximation in \eqref{eq:paprox} will be invalid. Our numerical studies indicate that so long as $c$ is relatively small, the performance of our method won't be affected much. Thus we set $c=0.01$ throughout the numerical studies in Section~\ref{sec:ns}.

\subsection{Permutation test}\label{sec:pt}

In the previous subsection, we mentioned when the correlation structure of the covariates is unknown, we can still obtain the $p$-value approximately using the proposed asymptotic distributions. In fact, the $p$-value can also be computed using the permutation test, which is a well-known resampling procedure that has many applications. A permutation test is applicable if the samples are exchangeable when the null hypothesis holds. In fact, under certain assumptions, the exchangeability condition can be satisfied.

\begin{remark}\label{thm:perm}
Suppose $\tilde{\by}$ is a random permuted sample from $\by$ and we obtain the test statistic as $\rpn(\tilde{\by}, \bX)$. If $Y$ is independent of the covariates, i.e., $\bbeta = \mathbf{0}$, then $\rpn(\tilde{\by}, \bX)$ has the same distribution as $\rpn$.
\end{remark}

To conduct the permutation test, at each step of the sequential selection, we randomly permute the observations of~$Y$ and obtain a new sample. Then we can compute the test statistic based on the new sample. The permutations are implemented repeatedly, and the $p$-value is obtained by the ranking of the original test statistic among the permuted test statistics over the total number of permutations. We further illustrate the permutation test step by step as below:

\begin{enumerate}
	\item
	At Step~$k$, we shuffle the observations of $Y$ at random $Q$ times and obtain the permuted sample $Y^{(q)} = (y^{q}_1, \ldots, y^q_n)$ for $q \in \{ 1,  \ldots, Q\}$.
	\item
	
	Compute the corresponding test statistic $\rpn^q$ for each $Y^{(q)}$, and compare the test statistic $\rpn$ obtained from the original $Y$.
	\item

	Suppose the rank of $\rpn$ among $\rpn^1, \ldots, \rpn^Q$ is $r_k$. Then the $p$-value of the permutation test can be written as $p_k = {r_k}/{Q}$.
\end{enumerate}

Recall that our goal is to use the distribution information to provide guidance for sequential selection procedures. In what follows, we introduce a test-based variable selection procedure by applying the results obtained in Section~\ref{sec:md}.

\section{Sequential testing for variable selection}\label{sec:seqtest}

\subsection{Testing-based variable selection procedure}\label{sec:proce}

For sequential selection procedures, it is crucial to find a stopping criterion. In other words, at each step of a particular selection procedure, we want to know whether there are remaining important covariates in the inactive set. Therefore, we propose to conduct the dependence test introduced in the previous section correspondingly at each step and stop the procedure once we accept the null hypothesis. This leads to a test-based variable selection approach.

Suppose we are at Step $k$  ($k\ge 1$) of a sequential selection procedure, and let $\cala_{k-1}$ denote the active set that includes the indices of selected variables from the previous step. We want to emphasize that here $\cala_{k-1}$ is fixed given the data. In contrast, we use the notation $\hat{\cala}_{k-1}(\bX, Y)$ to denote the index set for sampling from the data, which is random. Then one needs to know whether the remaining inactive covariates are all uncorrelated with the response, which is equivalent to testing \eqref{eq:null} with $\calm = \cala_{k-1}$ under the Gaussian assumption. Note that $\cala_0=\emptyset$ when $k=1$. More specifically,  we consider the following null hypothesis at Step $k$:
\begin{equation} \label{eq:null2}
\mathcal{H}_0^{(k)}: \text{ Conditioning on }X_{\cala_{k-1}},Y \text{ and } X_j \text{ are independent for }\forall j\notin \cala_{k-1}.
\end{equation}

We note here that the proposed testing in Section~\ref{sec:md} conditions on $X_{\cala_{k-1}}$, where $\cala_{k-1}$ is non-random, rather than on both $X_{\cala_{k-1}}$ and $\hat{\cala}_{k-1}(\bX, Y) = \cala_{k-1}$. However, below are a few justifications for using the proposed test in the model selection procedure.
\begin{enumerate}
  \item The main purpose of using the test in Section~\ref{sec:md} is to control the entry of variables with spurious partial correlation in the selection process. The ultimate goal is to assist the selected model in having good properties on FP, FN and MSE. In this regard, the problem is essentially different from post-selection inference \citep{fithian15, taylor14exact}, where the aim is to obtain valid conclusions for scientific discoveries. The simulation and real data studies in Section~\ref{sec:ns} demonstrate the good model selection properties of the proposed procedure.
  \item In the Online Supplement, we compare the empirical distributions of the unconditional test statistic in Section~\ref{sec:md} and the conditional ones through extensive simulations. We find that the difference is very small.
  \item The unconditional test provides a valid $p$-value at the first step of model selection to prevent any spurious variables from entering the model when $\bbeta = \mathbf{0}.$ For later steps, our test provides a good approximation of spurious correlation control.
\end{enumerate}
Based on the above considerations, we propose to incorporate the test in Section~\ref{sec:md} in the sequential selection procedure. The procedure is detailed below. Under \eqref{eq:null2}, the corresponding test statistic can be written as
\begin{equation}\label{eq:seqtest}
R^{(k)} = \max_{j:j\in\cala_{k-1}^c}|\hcorr(\rj^{(k)}, \br^{(k)})|,
\end{equation}
where
$$
\rj^{(k)} = (I-\proja)\bx_j, \quad \br^{(k)} = (I-\proja)\by
$$
with $\proja$ defined in the similar way as in Section \ref{sec:gtest}. Note that when $k = 1$, we have $\rj^{(1)}=\bx_j - \barxj$, where $\bar{x}_j$ is the mean of $\bx_j$ and $\onen$ is an $n$-dimensional  vector of $1$s since $P_{\cala_0} = \projone$. Similarly $\br^{(1)}$ reduces to $\by-\bar{y} \, \onen$.

From Theorem \ref{thm:corr}, we can see that when the covariates are independent, the $p$-value of $R^{(k)}$ converges to a uniform distribution on the unit interval, $\mathcal{U}(0,1)$, under null hypothesis \eqref{eq:null2}. This conclusion is formally stated below.

\begin{corollary}\label{thm:puni}
	Suppose we have a linear model as in \eqref{eq:lm} and we assume that the covariates are independent Gaussian variables. Let $x^{(k)}$ be the observed value of the test statistic $R^{(k)}$ as defined in~\eqref{eq:seqtest}. Then the $p$-value can be obtained from $p(x^{(k)}) = 1-F_{n,k-1}(x^{(k)})$. Under the null hypothesis~\eqref{eq:null2}, we have $p(x^{(k)})\limd \mathcal{U}(0,1)$ as $\infp$.
\end{corollary}

We omit the proof because it follows directly from Theorem \ref{thm:corr}. Corollary \ref{thm:puni} suggests that it is possible and reasonable to use the proposed test statistic $R^{(k)}$ when the covariates are independent Gaussian variables. For dependent covariates, although we do not have similar theoretical results for the distribution of the $p$-value, we can use the approximation described in Section \ref{sec:cor} to obtain the $p$-value. Our numerical studies demonstrate that such an approximation can work well.

Thus far we have discussed how to construct our dependency tests sequentially. Now we introduce our test-based variable selection method. In each step of the selection procedure, we compute the current test statistic and the corresponding $p$-value, and stop the selection when the $p$-value exceeds a pre-defined level $\gamma$. More specifically, our method is implemented in the following way.
\begin{enumerate}
	\item
	Set the active set to be $\cala_0=\emptyset$.
	\item
	\begin{enumerate}
		\item
		In the $k$th step ($k\ge 1$), compute the residuals $\br_j^{(k)} = (I-\proja)\bx_j$ and $\br^{(k)} = (I-\proja)\by$ for each inactive covariate $X_j$ and the response, respectively. Then derive the test statistic $R^{(k)}$ as in~\eqref{eq:seqtest}.
		\item
		Compute the $p$-value $p_k$ as in \eqref{eq:piid} for independent covariates and \eqref{eq:pcor} for dependent covariates.
	\end{enumerate}
	\item
	If $p_k\le\gamma$ and $k\le n-2$, update the active set $\cala_{k}$ and get the estimates of $\bbeta$ using the same approach as the original selection procedure; otherwise, terminate the procedure.
\end{enumerate}

In the above procedure, the stopping criterion in Step 3 involves a constant level $\gamma$. Here we do not provide a specific value of $\gamma$, because the choice of an appropriate $\gamma$ should depend on the goal of the selection, which might vary in different contexts. More specifically, if we aim to detect important variables other than losing any information, we could set a large $\gamma$. However, if we want to avoid false discoveries, we should choose a small $\gamma$. We will illustrate the effect of $\gamma$ by simulation examples in Section~\ref{sec:ns}. In practice, we also need to determine which null distribution to use in order to obtain the $p$-value. As mentioned in Section~\ref{sec:cor}, we first compute the average of the pairwise sample correlation among the covariates, say $\hat{\rho}$, to estimate $\rho$. If $|\hat{\rho}|<0.01$, we use~\eqref{eq:piid} to compute the $p$-value; otherwise we apply \eqref{eq:pcor} instead.

 Our method conducts a sequence of hypothesis tests adaptively until the null hypothesis \eqref{eq:null2} is accepted. Moreover, at each step we perform the dependency test before adding the next variable into the active set, which stands alone from the original variable selection procedure. Hence the proposed method essentially adds (or drops in the LASSO path) the variables one by one in the same order as in the original sequential selection approach. This property makes our method very flexible because it can be incorporated into any sequential selection procedure.

\subsection{Prostate cancer data example}\label{sec:toyeg}

In Section~\ref{sec:proce}, we have discussed how to implement our test-based variable selection approach in sequential selection procedures. To better illustrate how our method works, we apply it to the prostate cancer data, which has been well studied in the literature \citep{tib96}. This dataset contains 97 observations and eight predictor variables, of which 67 are training samples. The study goal is to predict the logarithm of prostate-specific antigen level (\emph{lpsa}) of men who were about to receive a radical prostatectomy.

We incorporate our approach into LARS and perform the variable selection on the training data. At each LARS step, we obtain the variable that enters into the model, the corresponding active set as well as the $p$-value. As the average of pairwise correlation is about 0.3, we use \eqref{eq:pcor} to compute the $p$-value. The results are reported in Table~\ref{t:pros}. It must be pointed out that the $p$-value is not associated with each variable, but the inactive set $\cala_{k-1}^c$ at each selection step. For example, the $p$-value $0.0010$ at Step~1 means that given the selected variable \emph{lcavol}, there is strong evidence that there is at least one important variable in the inactive set $\cala_1^c$. If one sets the constant level $\gamma$ described in Section~\ref{sec:proce} to be $0.1$, the selected variables are \emph{lcavol}, \emph{lweight} and \emph{svi}; if $\gamma$ is increased to $0.5$, there is one more variable \emph{lbph} added into the final model.

\begin{table}[!t]
	\centering
	\caption{Testing-based LARS procedure applied to the prostate cancer data. For each step, we report the variable selected by LARS, the active set $\cala_{k-1}$ in null hypothesis \eqref{eq:null2} and the $p$-value obtained from our testing approach. The stepwise $p$-value is calculated before the selected variable enters the candidate model.}
	\label{t:pros}
	\begin{tabular}{llll}
		\noalign{\vskip 5pt}
		Step & Variable Selected & Active Set $\cala_{k}$ & $p$-value  \\\hline
		0    &         & $\emptyset$ & 0.0000 \\
		1    & lcavol  & {1}  & 0.0010 \\
		2    & lweight & {1, 2} & 0.0791 \\
		3    & svi     & {1, 2, 5} &0.0645 \\
		4    & lbph    & {1, 2, 5, 4} & 0.2996 \\
		5    & pgg45   & {1, 2, 5, 4, 8} & 0.9482 \\
		6    & age     & {1, 2, 5, 4, 8, 3} & 0.7591 \\
		7    & lcp     & {1, 2, 5, 4, 8, 3, 6}& 0.5681 \\
		8    & gleason & {1, 2, 5, 4, 8, 3, 6, 7} & \\\hline
	\end{tabular}
\end{table}

\section{Numerical studies}\label{sec:ns}

In this section, we explore the performance of our method in terms of both simulation and real data studies. We incorporate the proposed approach into sequential selection procedures and compare the results with that using 10-fold CV to conduct model selection for each particular procedure.

\subsection{Simulation study}\label{sec:sim}

In our simulation experiments, we consider three sequential selection procedures: LARS, LASSO and FSR. When our test-based approach is incorporated into a particular procedure, we denote the corresponding variable selection method as LARS-Corr. Similarly we use the notations LASSO-Corr and FSR-Corr to represent our methods integrated with LASSO and FSR, respectively. In addition, we perform permutation tests in each of these three variable selection procedures and denote the corresponding methods by LARS-Perm, LASSO-Perm and FSR-Perm, respectively. For comparison, we use 10-fold CV in LARS, LASSO and FSR to implement model selection. We represent these three CV-based methods by LARS-CV, LASSO-CV and FSR-CV. We also perform the truncated Gaussian tests in the sequential selection procedures LARS and FSR, denoted as LARS-TG, FSR-TG, respectively. For permutation tests, we implement 500 permutations. Due to space limit, we only present the results for LARS and LASSO here, while the results for FSR are shown in the Online Supplement since they lead to similar conclusions.

\begin{table*}[t!]
	\centering
	\def\arraystretch{1.05}
	\caption{Results for simulated Example 1 with $\rho=0$ and $\sigma=2$. For each method, we report the average MSE, FN, FP and computational time over 100 replications (with standard errors given in parentheses). For our approaches, we show the results with $\gamma = 0.01, 0.05, 0.2, 0.5$ in the stopping criterion described in Section \ref{sec:proce}. For each sequential selection procedure, we highlight the smallest MSE and run time in bold font. One can see that the performance of the proposed method is competitive to CV and is more computationally efficient.}
	\label{t:eg11}
\scriptsize
	\begin{tabular}{llrrrr}
		Methods & $\gamma$  & \multicolumn{1}{c}{MSE} & \multicolumn{1}{c}{FN}  & \multicolumn{1}{c}{FP}  & \multicolumn{1}{c}{Time}  \\\hline
		\noalign{\vskip 3pt}
		LARS-CV   &     & 4.35 (0.05)  & 0.00 (0.00) & 1.85 (0.32)  & 28.37 (0.15)  \\ [2pt]
		LARS-Perm & 0.01 & {\bf 4.05 (0.03)} & 0.00 (0.00) & 0.01 (0.01) & 5.70 (0.04) \\
		LARS-Perm & 0.05 & 4.08 (0.03) & 0.00 (0.00) & 0.10 (0.04) & 5.82 (0.07) \\
		LARS-Perm &  0.2 & 4.15 (0.04) & 0.00 (0.00) & 0.40 (0.08) & 6.29 (0.13) \\
		LARS-Perm & 0.5 & 4.36 (0.05) & 0.00 (0.00) & 2.04 (0.41) & 8.70 (0.58)        \\ [2pt]
		LARS-Corr & 0.01 & {\bf 4.05 (0.03)} & 0.00 (0.00) & 0.00 (0.00) & 0.76 (0.02) \\
		LARS-Corr & 0.05 & 4.07 (0.03) & 0.00 (0.00) & 0.08 (0.03) & {\bf  0.70 (0.01)} \\
		LARS-Corr & 0.2 & 4.13 (0.04) & 0.00 (0.00) & 0.32 (0.08) & 0.83 (0.02) \\	
		LARS-Corr & 0.5 & 4.33 (0.05) & 0.00 (0.00) & 1.44 (0.22) & 1.03 (0.05) \\ [2pt]
		LARS-TG & 0.01 & 10.22 (0.08) & 1.99 (0.01) & 0.00 (0.00) & 12.16 (0.12) \\
		LARS-TG & 0.05 & 9.89 (0.13) & 1.90 (0.03) & 0.00 (0.00) & 12.17 (0.12)\\
		LARS-TG & 0.2 & 8.89 (0.23) & 1.62 (0.06) & 0.01 (0.01) & 12.41 (0.14)\\
		LARS-TG & 0.5 & 6.85 (0.26) & 0.97 (0.08) & 0.23 (0.06) & 12.31 (0.13)\\
		[6pt]
		LASSO-CV  &    & 4.70 (0.06) & 0.00 (0.00) & 4.78 (0.70) & 39.74 (0.58) \\ [2pt]
		LASSO-Perm & 0.01 & 4.07 (0.03) & 0.00 (0.00) & 0.00 (0.00) & 5.67 (0.04) \\
		LASSO-Perm & 0.05 & 4.08 (0.03) & 0.00 (0.00) & 0.03 (0.02) & 5.72 (0.06) \\
		LASSO-Perm & 0.2 & 4.17 (0.04) & 0.00 (0.00) & 0.40 (0.09) & 6.31 (0.14) \\
		LASSO-Perm & 0.5 & 4.36 (0.05) & 0.00 (0.00) & 1.76 (0.32) & 8.35 (0.45)	\\	 [2pt]
		LASSO-Corr & 0.01 &  {\bf 4.07 (0.03)} & 0.00 (0.00) & 0.00 (0.00) &  0.70 (0.01) \\
		LASSO-Corr& 0.05  & 4.08 (0.03) & 0.00 (0.00) & 0.02 (0.01)& { \bf 0.70 (0.00)} \\
		LASSO-Corr& 0.2  & 4.13 (0.03) & 0.00 (0.00) & 0.25 (0.06) & 0.83 (0.02)\\
		LASSO-Corr& 0.5 & 4.34 (0.04) & 0.00 (0.00) & 1.46 (0.24) & 1.07 (0.07) \\
		\hline		         	
	\end{tabular}
\end{table*}

\begin{table*}[t!]
\scriptsize
	\centering
	\def\arraystretch{1.05}
	\caption{Results for simulated Example 1 with $\rho=0$  and $\sigma=6$. The format of the table is the same as Table \ref{t:eg11}. In general, the performance of the proposed method is competitive to CV and is more computationally efficient.}
	\label{t:eg11large}
	\begin{tabular}{llrrrr}
		Methods & $\gamma$  & \multicolumn{1}{c}{MSE} & \multicolumn{1}{c}{FN}  & \multicolumn{1}{c}{FP}  & \multicolumn{1}{c}{Time}  \\\hline
		\noalign{\vskip 3pt}
		LARS-CV   &      & 41.33 (0.48) & 1.24 (0.09) & 0.82 (0.23)  & 27.31 (0.14)  \\[2pt]
		LARS-Perm & 0.01 & 40.33 (0.39) & 1.40 (0.07) & 0.03 (0.02) & 3.88 (0.11) \\
		LARS-Perm & 0.05 & 40.05 (0.40) & 1.25 (0.07) & 0.14 (0.04) & 4.23 (0.13) \\
		LARS-Perm & 0.2 & {\bf 39.88 (0.41)} & 1.01 (0.06) & 0.48 (0.08) & 5.06 (0.17) \\
		LARS-Perm & 0.5 & 41.31 (0.49) & 0.75 (0.06) & 1.99 (0.33) & 7.66 (0.50) 	\\ [2pt]	
		LARS-Corr & 0.01 & 40.37 (0.38) & 1.42 (0.07) & 0.02 (0.01) & {\bf 0.44 (0.02)} \\
		LARS-Corr & 0.05 & 40.10 (0.39) & 1.27 (0.07) & 0.13 (0.04) & 0.47 (0.02) \\
		LARS-Corr & 0.2  &  39.90 (0.41) & 1.02 (0.06) & 0.47 (0.09) & 0.61 (0.03)\\	
		LARS-Corr & 0.5  & 41.26 (0.48) & 0.78 (0.06) & 1.60 (0.20) & 0.97 (0.06)\\[2pt]
		LARS-TG & 0.01 & 43.57 (0.36) & 2.11 (0.03) & 0.01 (0.01) & 12.86 (0.22)\\
		LARS-TG & 0.05 & 43.26 (0.34) & 2.06 (0.03) & 0.02 (0.02) & 13.07 (0.23)\\
		LARS-TG & 0.2 & 42.48 (0.32) & 1.91 (0.04) & 0.05 (0.03) & 13.03 (0.22)\\
		LARS-TG & 0.5 & 41.70 (0.36) & 1.53 (0.06) & 0.43 (0.08) & 13.11 (0.22)\\
		[6pt]
		LASSO-CV  &     & 42.33 (0.54) & 1.16 (0.08) & 1.94 (0.60) & 35.52 (0.24) \\[2pt]
		LASSO-Perm & 0.01 & 41.21 (0.38) & 1.56 (0.06) & 0.01 (0.01) & 3.58 (0.11)\\
		LASSO-Perm & 0.05 & 40.34 (0.36) & 1.28 (0.06) & 0.09 (0.04) & 4.14 (0.13)\\
		LASSO-Perm & 0.2 & 40.10 (0.38) & 1.02 (0.06) & 0.43 (0.08) & 4.97 (0.17)\\
		LASSO-Perm & 0.5 & 41.46 (0.45) & 0.76 (0.07) & 1.73 (0.24) & 7.23 (0.39)\\ [2pt]
		LASSO-Corr& 0.01 & 41.30 (0.38) & 1.58 (0.06) & 0.01 (0.01) & {\bf 0.39 (0.01)} \\
		LASSO-Corr& 0.05 & 40.36 (0.36) & 1.30 (0.06) & 0.06 (0.03) & 0.47 (0.02) \\
		LASSO-Corr& 0.2  & {\bf 40.00 (0.38)} & 1.02 (0.06) & 0.39 (0.08) & 0.61 (0.02)\\
		LASSO-Corr& 0.5 & 41.41 (0.44) & 0.80 (0.06) & 1.49 (0.18) & 0.88 (0.04) \\
		\hline		       	
	\end{tabular}
\end{table*}

\begin{table*}[t!]
	\scriptsize
	\centering
	\def\arraystretch{1.05}
	\caption{Results for simulated Example 1 with $\rho =0.3$ and $\sigma=2$. The format of the table is the same as Table \ref{t:eg11}. In general, the performance of the proposed method is competitive to CV and is more computationally efficient.}
	\label{t:eg12}
	\begin{tabular}{llrrrr}
		Methods & $\gamma$ & \multicolumn{1}{c}{MSE} & \multicolumn{1}{c}{FN}  & \multicolumn{1}{c}{FP}  & \multicolumn{1}{c}{Time}  \\ \hline
		\noalign{\vskip 3pt}	
		LARS-CV   &      & 4.52 (0.06) & 0.00 (0.00) & 4.72 (0.86) & 28.31 (0.24)  \\ [2pt]
		LARS-Perm & 0.01 & {\bf 4.06 (0.03)} & 0.00 (0.00) & 0.04 (0.02) & 5.23 (0.04) \\
		LARS-Perm & 0.05 & 4.07 (0.03) & 0.00 (0.00) & 0.07 (0.03) & 5.26 (0.05) \\
		LARS-Perm & 0.2 & 4.13 (0.04) & 0.00 (0.00) & 0.33 (0.09) & 5.62 (0.12) \\
		LARS-Perm & 0.5 & 4.51 (0.09) & 0.00 (0.00) & 7.64 (2.81) & 15.61 (3.85) \\ [2pt]
		LARS-Corr & 0.01 & 4.08 (0.03) & 0.00 (0.00) & 0.10 (0.03) & {\bf 0.54 (0.01)} \\
		LARS-Corr & 0.05 & 4.09 (0.03) & 0.00 (0.00) & 0.16 (0.04) & 0.55 (0.01)\\
		LARS-Corr & 0.2  & 4.14 (0.03) & 0.00 (0.00) & 0.47 (0.08) & 0.57 (0.01) \\	
		LARS-Corr & 0.5  & 4.29 (0.04) & 0.00 (0.00) & 1.84 (0.36) & 0.76 (0.04) \\ [2pt]
		LARS-TG & 0.01 & 8.46 (0.05) & 2.00 (0.00) & 0.00 (0.00) & 10.97 (0.03)\\
		LARS-TG & 0.05 & 8.33 (0.07) & 1.95 (0.02) & 0.00 (0.00) & 11.02 (0.04)\\
		LARS-TG & 0.2 & 7.79 (0.12) & 1.72 (0.05) & 0.00 (0.00) & 11.12 (0.05)\\
		LARS-TG & 0.5 & 6.95 (0.16) & 1.35 (0.07) & 0.05 (0.03) & 11.09 (0.04)\\
		[6pt]
		LASSO-CV  &     & 4.73 (0.06) & 0.00 (0.00) & 6.69 (0.83) & 38.05 (0.53) \\ [2pt]
		LASSO-Perm & 0.01 & {\bf 4.07 (0.03)} & 0.00 (0.00) & 0.00 (0.00) & 5.67 (0.04) \\
		LASSO-Perm & 0.05 & 4.08 (0.03) & 0.00 (0.00) & 0.03 (0.02) & 5.72 (0.06)\\
		LASSO-Perm & 0.2 & 4.17 (0.04) & 0.00 (0.00) & 0.40 (0.09) & 6.31 (0.14)\\
		LASSO-Perm & 0.5 & 4.36 (0.05) & 0.00 (0.00) & 1.76 (0.32) & 8.35 (0.45)\\ [2pt]
		LASSO-Corr & 0.01 &  4.11 (0.03) & 0.00 (0.00) & 0.09 (0.03) & {\bf 0.55 (0.01)} \\
		LASSO-Corr& 0.05 & 4.12 (0.03) & 0.00 (0.00) & 0.15 (0.04) & 0.56 (0.01) \\
		LASSO-Corr& 0.2 & 4.17 (0.03) & 0.00 (0.00) & 0.37 (0.06) & 0.59 (0.01)\\
		LASSO-Corr& 0.5 & 4.30 (0.04) & 0.00 (0.00) & 1.58 (0.32) & 0.76 (0.03) \\	
		\hline
	\end{tabular}
\end{table*}		

\begin{table*}[t!]
\scriptsize
	\centering
	\def\arraystretch{1.05}
	\caption{Results for simulated Example 1 with $\rho =0.3$ and $\sigma=6$. The format of the table is the same as Table \ref{t:eg11}. In general, the performance of the proposed method is competitive to CV and is more computationally efficient.}
	\label{t:eg12large}
	\begin{tabular}{llrrrr}
		Methods & $\gamma$ & \multicolumn{1}{c}{MSE} & \multicolumn{1}{c}{FN}  & \multicolumn{1}{c}{FP}  & \multicolumn{1}{c}{Time}  \\ \hline
		\noalign{\vskip 3pt}	
		LARS-CV   &      & 41.89 (0.51) & 1.38 (0.07) & 2.40 (0.76) & 29.13 (0.24)  \\ [2pt]
		LARS-Perm & 0.01 & 40.66 (0.31) & 1.88 (0.04) & 0.02 (0.01) & 2.65 (0.06)  \\
		LARS-Perm & 0.05 & 40.59 (0.37) & 1.70 (0.06) & 0.31 (0.13) & 3.25 (0.21)  \\
		LARS-Perm & 0.2 & 42.83 (0.57) & 1.33 (0.08) & 5.82 (2.05) & 10.89 (2.67)  \\
		LARS-Perm & 0.5 & 48.02 (0.94) & 0.90 (0.08) & 25.78 (5.17) & 37.59 (6.84) \\ [2pt]
		LARS-Corr & 0.01 & 40.61 (0.31) & 1.77 (0.05) & 0.01 (0.01) & {\bf 0.26 (0.02)} \\
		LARS-Corr & 0.05 & 40.34 (0.30) & 1.62 (0.06) & 0.10 (0.03) & 0.32 (0.03)\\
		LARS-Corr & 0.2  & {\bf 39.95 (0.33)} & 1.41 (0.06) & 0.23 (0.05) & 0.44 (0.05)\\	
		LARS-Corr & 0.5  & 40.39 (0.37) & 1.25 (0.06) & 1.03 (0.28) & 0.56 (0.05) \\[2pt]
		LARS-TG & 0.01 & 42.21 (0.42) & 2.11 (0.03) & 0.00 (0.00) & 11.47 (0.06)\\
		LARS-TG & 0.05 & 41.73 (0.37) & 2.03 (0.03) & 0.09 (0.06) & 11.46 (0.06)\\
		LARS-TG & 0.2 & 41.46 (0.37) & 1.91 (0.04) & 0.29 (0.09) & 11.46 (0.06)\\
		LARS-TG & 0.5 & 41.45 (0.35) & 1.66 (0.05) & 1.22 (0.20) & 11.82 (0.08)\\
		[6pt]
		LASSO-CV  &     & 42.26 (0.54) & 1.36 (0.07) & 2.58 (0.73) & 39.07 (0.56) \\[2pt]
		LASSO-Perm & 0.01 & 41.21 (0.38) & 1.56 (0.06) & 0.01 (0.01) & 3.58 (0.11) \\
		LASSO-Perm & 0.05 & 40.34 (0.36) & 1.28 (0.06) & 0.09 (0.04) & 4.14 (0.13)\\
		LASSO-Perm & 0.2 & {\bf 40.10 (0.38)} & 1.02 (0.06) & 0.43 (0.08) & 4.97 (0.17)\\
		LASSO-Perm & 0.5 & 41.46 (0.45) & 0.76 (0.07) & 1.73 (0.24) & 7.23 (0.39)\\
		LASSO-Corr& 0.01 & 41.38 (0.39) & 1.82 (0.05) & 0.22 (0.15) & {\bf 0.23 (0.01)} \\
		LASSO-Corr& 0.05 & 40.90 (0.40) & 1.60 (0.06) & 0.36 (0.17) & 0.28 (0.02)\\
		LASSO-Corr& 0.2 &  40.62 (0.40) & 1.45 (0.06) & 0.46 (0.17) & 0.43 (0.04) \\
		LASSO-Corr& 0.5 & 40.75 (0.43) & 1.25 (0.06) & 1.18 (0.32) & 0.55 (0.05)\\
		\hline	
	\end{tabular}
\end{table*}

\begin{table*}[t!]
\scriptsize
	\centering
	\def\arraystretch{1.05}
	\caption{Results for simulated Example 2. The format of the table is the same as Table \ref{t:eg11}. One can see that the performance of the proposed method is competitive to CV and is more computationally efficient.}
	\label{t:eg2}
	\begin{tabular}{llrrrr}
		Methods & $\gamma$ & \multicolumn{1}{c}{MSE} & \multicolumn{1}{c}{FN}  & \multicolumn{1}{c}{FP}  & \multicolumn{1}{c}{Time}  \\\hline
		\noalign{\vskip 3pt}	
		LARS-CV   &      & 10.78 (0.14) & 0.00 (0.00) & 4.04 (0.50) & 27.25 (0.19)  \\ [2pt]
		LARS-Perm & 0.01 & 9.57 (0.09) & 0.04 (0.02) & 0.04 (0.02) & 14.76 (0.11) \\
		LARS-Perm & 0.05 & 9.61 (0.09) & 0.02 (0.01) & 0.18 (0.07) & 14.95 (0.15) \\
		LARS-Perm & 0.2 & 9.85 (0.11) & 0.01 (0.01) & 0.59 (0.12) & 15.69 (0.19) \\
		LARS-Perm & 0.5 & 10.51 (0.13) & 0.01 (0.01) & 2.60 (0.36) & 18.24 (0.52) \\ [2pt]
		LARS-Corr & 0.01 & 9.56 (0.09) & 0.02 (0.01) & 0.11 (0.06) & {\bf 1.73 (0.01)} \\
		LARS-Corr & 0.05 & {\bf 9.55 (0.08)} & 0.01 (0.01) & 0.12 (0.07) & 1.74 (0.02)\\
		LARS-Corr & 0.2 & 9.77 (0.09) & 0.01 (0.01) & 0.52 (0.12) & 1.80 (0.03)\\
		LARS-Corr & 0.5 & 10.25 (0.13) & 0.01 (0.01) & 3.67 (1.88) & 2.40 (0.37)\\ [2pt]
		LARS-TG & 0.01 & 12.36 (0.14) & 1.98 (0.02) & 0.00 (0.00) & 12.38 (0.15)\\
		LARS-TG & 0.05 & 12.32 (0.14) & 1.95 (0.03) & 0.02 (0.01) & 12.24 (0.13)\\
		LARS-TG & 0.2 & 11.83 (0.10) & 1.75 (0.04) & 0.10 (0.04) & 12.41 (0.13)\\
		LARS-TG & 0.5 & 11.51 (0.11) & 1.48 (0.05) & 0.45 (0.10) & 12.46 (0.14)\\
		[6pt]
		LASSO-CV  &     &  12.02 (0.12) & 0.00 (0.00) & 10.41 (0.84) & 40.09 (0.30) \\ [2pt]
		LASSO-Perm & 0.01 & {\bf 9.57 (0.08)} & 0.02 (0.01) & 0.03 (0.02) & 14.68 (0.09) \\
		LASSO-Perm & 0.05 & 9.61 (0.08) & 0.01 (0.01) & 0.14 (0.07) & 14.86 (0.12)\\
		LASSO-Perm & 0.2 & 10.02 (0.13) & 0.01 (0.01) & 1.25 (0.45) & 16.60 (0.76)\\
		LASSO-Perm & 0.5 & 10.75 (0.15) & 0.01 (0.01) & 3.59 (0.61) & 19.33 (0.78)\\ [2pt]
		LASSO-Corr & 0.01 & {\bf 9.57 (0.08)} & 0.01 (0.01) & 0.09 (0.06) & {\bf 1.72 (0.01)} \\
		LASSO-Corr & 0.05 & 9.65 (0.09) & 0.01 (0.01) & 0.33 (0.17) & 1.76 (0.03)\\
		LASSO-Corr & 0.2 & 9.90 (0.11) & 0.01 (0.01) & 1.01 (0.43) & 1.87 (0.08)\\
		LASSO-Corr & 0.5 & 10.4 (0.14) & 0.01 (0.01) & 2.56 (0.51) & 2.17 (0.09)\\	
		\hline		
	\end{tabular}
\end{table*}

\begin{table*}[t!]
	\scriptsize
	\centering
	\def\arraystretch{1.05}
	\caption{Results for simulated Example 3 with $\sigma = 4$. The format of the table is the same as Table \ref{t:eg11}. One can see that the performance of the proposed method is competitive to CV and is more computationally efficient.}
	\label{t:eg3}
	\begin{tabular}{llrrrr}
		Methods & $\gamma$ & \multicolumn{1}{c}{MSE} & \multicolumn{1}{c}{FN}  & \multicolumn{1}{c}{FP}  & \multicolumn{1}{c}{Time}  \\\hline
		\noalign{\vskip 3pt}	
		LARS-CV   &      & 17.65 (0.21) & 0.00 (0.00) & 2.15 (0.37) & 26.57 (0.23) \\ [2pt]
		LARS-Perm & 0.01 & 16.25 (0.13) & 0.02 (0.01) & 0.00 (0.00) & 5.74 (0.05)  \\
		LARS-Perm & 0.05 & 16.28 (0.13) & 0.01 (0.01) & 0.04 (0.03) & 5.81 (0.07) \\
		LARS-Perm & 0.2 & 16.64 (0.15) & 0.00 (0.00) & 0.43 (0.10) & 6.26 (0.16) \\
		LARS-Perm & 0.5 & 17.49 (0.24) & 0.00 (0.00) & 3.58 (1.96) & 10.65 (2.68) \\ [2pt]
		LARS-Corr & 0.01 & {\bf 16.25 (0.13)} & 0.02 (0.01) & 0.00 (0.00) & 0.93 (0.02) \\
		LARS-Corr & 0.05 & 16.28 (0.13) & 0.01 (0.01) & 0.04 (0.03) & {\bf 0.88 (0.01)} \\
		LARS-Corr & 0.2 & 16.61 (0.15) &  0.00 (0.00) & 0.39 (0.09) & 1.04 (0.04) \\
		LARS-Corr & 0.5 & 17.26 (0.19) & 0.00 (0.00) & 1.27 (0.21) & 1.21 (0.06) \\ [2pt]
		LARS-TG & 0.01 & 38.07 (0.57) & 2.00 (0.03) & 0.00 (0.00) & 12.00 (0.06)\\
		LARS-TG & 0.05 & 37.71 (0.56) & 1.95 (0.04) & 0.02 (0.02) & 12.02 (0.06)\\
		LARS-TG & 0.2 & 36.69 (0.50) & 1.80 (0.05) & 0.04 (0.03) & 11.95 (0.05)\\
		LARS-TG & 0.5 & 34.24 (0.57) & 1.24 (0.08) & 0.51 (0.14) & 12.30 (0.09)\\
		[6pt]
		LASSO-CV  &     & 18.16 (0.24) & 0.00 (0.00) & 3.27 (0.49) & 44.60 (0.51) \\ [2pt]
		LASSO-Perm & 0.01 & 16.44 (0.13) & 0.03 (0.02) & 0.02 (0.01) & 5.70 (0.05) \\
		LASSO-Perm & 0.05 & 16.41 (0.11) & 0.01 (0.01) & 0.06 (0.02) & 5.77 (0.06)\\
		LASSO-Perm & 0.2 & 16.65 (0.14) & 0.00 (0.00) & 0.34 (0.09) & 6.10 (0.13)\\
		LASSO-Perm & 0.5 & 17.42 (0.21) & 0.00 (0.00) & 3.21 (1.94) & 11.66 (4.35)\\ [2pt]
		LASSO-Corr & 0.01 & {\bf 16.37 (0.12)} & 0.02 (0.01) & 0.00 (0.00) & 1.07 (0.02) \\
		LASSO-Corr & 0.05 &	16.41 (0.11) & 0.01 (0.01) & 0.05 (0.02) & {\bf 1.00 (0.02)} \\
		LASSO-Corr & 0.2 &	16.64 (0.14) & 0.00 (0.00) & 0.33 (0.09) & 1.03 (0.03) \\
		LASSO-Corr & 0.5 &	17.22 (0.17) & 0.00 (0.00) & 1.08 (0.19) & 1.08 (0.04) \\
		\hline		
	\end{tabular}
\end{table*}

\begin{table*}[b!]
		\scriptsize
	\centering
	\def\arraystretch{1.05}
	\caption{Results for simulated Example 3 with $\sigma = 8$. The format of the table is the same as Table \ref{t:eg11}. One can see that the performance of the proposed method is competitive to CV and is more computationally efficient.}
	\label{t:eg3large}
	\begin{tabular}{llrrrr}
		Methods & $\gamma$ & \multicolumn{1}{c}{MSE} & \multicolumn{1}{c}{FN}  & \multicolumn{1}{c}{FP}  & \multicolumn{1}{c}{Time}  \\\hline
		\noalign{\vskip 3pt}
		LARS-CV   &      & 76.59 (0.97) & 1.62 (0.10) & 0.97 (0.30) & 25.94 (0.08) \\ [2pt]
		LARS-Perm & 0.01 & 73.71 (0.71) & 1.78 (0.06) & 0.00 (0.00) & 3.03 (0.09) \\
		LARS-Perm & 0.05 & 72.04 (0.74) & 1.43 (0.07) & 0.07 (0.04) & 3.62 (0.12) \\
		LARS-Perm & 0.2 & {\bf 71.75 (0.77)} & 1.15 (0.07) & 0.41 (0.10) & 4.47 (0.18) \\
		LARS-Perm & 0.5 & 74.19 (0.94) & 0.93 (0.07) & 3.39 (1.95) & 8.68 (2.54) \\ [2pt]
		LARS-Corr & 0.01 &  73.72 (0.76) & 1.76 (0.06) & 0.01 (0.01) & {\bf 0.43 (0.02)} \\
		LARS-Corr & 0.05 & 72.23 (0.74) & 1.50 (0.07) & 0.03 (0.02) & 0.55 (0.02) \\
		LARS-Corr & 0.2 & 71.99 (0.78) & 1.19 (0.07) & 0.38 (0.09) & 0.71 (0.03) \\
		LARS-Corr & 0.5 &	73.61 (0.80) & 0.93 (0.07) & 1.22 (0.15) & 1.07 (0.05) \\[2pt]
		LARS-TG & 0.01 & 125.93 (1.84) & 2.47 (0.05) & 0.01 (0.01) & 12.11 (0.08) \\
		LARS-TG & 0.05 & 124.84 (1.82) & 2.35 (0.05) & 0.04 (0.02) & 12.26 (0.10) \\
		LARS-TG & 0.2 & 124.42 (1.78) & 2.25 (0.05) & 0.15 (0.05) & 12.26 (0.10) \\
		LARS-TG & 0.5 & 124.51 (1.72) & 2.00 (0.06) & 0.55 (0.10) & 12.29 (0.09) \\
		[6pt]
		LASSO-CV  &     & 72.70 (1.26) & 1.49 (0.11) & 1.06 (0.42) & 40.59 (0.30) \\ [2pt]
		LASSO-Perm & 0.01 & 73.83 (0.76) & 1.71 (0.07) & 0.01 (0.01) & 3.05 (0.10) \\
		LASSO-Perm & 0.05 & 72.33 (0.72) & 1.43 (0.08) & 0.07 (0.03) & 3.53 (0.12)\\
		LASSO-Perm & 0.2 & 72.21 (0.72) & 1.19 (0.08) & 0.34 (0.08) & 4.22 (0.18)\\
		LASSO-Perm & 0.5 & 74.24 (0.85) & 0.95 (0.07) & 3.14 (1.93) & 10.46 (4.80)\\ [2pt]
		LASSO-Corr& 0.01 & 73.91 (0.79) & 1.71 (0.07) & 0.02 (0.01) & {\bf 0.51 (0.02)} \\
		LASSO-Corr& 0.05 & 72.60 (0.74) & 1.49 (0.07) & 0.03 (0.02) & 0.53 (0.02) \\
		LASSO-Corr& 0.2 &	{\bf 72.09 (0.72)} & 1.19 (0.08) & 0.31 (0.08) & 0.71 (0.03) \\
		LASSO-Corr& 0.5 &	73.83 (0.77) & 0.98 (0.07) & 1.18 (0.16) & 1.01 (0.05) \\
		[6pt]		
		\hline		
	\end{tabular}
\end{table*}

\newpage

Let $\hat{\bbeta} = (\hat{\beta_1}, \ldots, \hat{\beta_p})^\top$ denote the estimated coefficient vector. We evaluate the variable selection accuracy by two quantities: False Negatives (FN) and False Positives (FP), respectively defined as
$$
FN = \sum_{j=1}^p \mathbf{1}(\hbeta_j=0)\times \mathbf{1} (\beta_j\ne0) \quad \mbox{and} \quad
FP = \sum_{j=1}^p \mathbf{1} (\hbeta_j\ne0)\times \mathbf{1} (\beta_j=0),
$$
where $\mathbf{1}$ denotes an indicator function.

We consider three simulated examples to generate the response variable. For the first two examples, the covariate vector $\bX$ is generated from a $p$-dimensional Gaussian distribution $\caln(0, \Sigma)$ with correlation matrix $\Sigma= (\rho_{i,j})$. For the third example, we aim to assess the robustness of our procedure, and therefore we generate independent covariates and random noise from a central Student $t$ distribution with 5~degrees of freedom. Throughout the simulation experiments, we fix $p=2000$. We generate 100 simulated datasets with $n=200$ observations from each model. In each replication, given a set of selected variables, we refit a linear model and calculate the out-of-sample mean squared errors (MSE) using an independent test dataset with $500$ observations. The details of the simulation examples are as follows.

\newpage

\bigskip
\noindent
{\bf Example 1}. We generate the response from the following sparse linear model
$Y =  3X_1-1.5X_2+2X_3+\veps$,
where the covariates have equal pairwise correlation, i.e., $\rho_{i,j} = \tcorr(X_i, X_j) = \rho$ for all $i\ne j$. We set $\rho = 0$ for independent covariates and $\rho = 0.3$ for dependent covariates. We also consider $\sigma = 2$ for strong signal and $\sigma = 6$ for weak signal.

\bigskip
\noindent
{\bf Example 2.} We demonstrate that when the covariates do not have equal pairwise correlations, we can still apply our approach using the approximated null distribution discussed in Section~\ref{sec:cor}. We simulate data from
$Y = 2X_1+\cdots+2X_{10}+\veps$,
where $\rho_{i,j} = 0.5^{|i-j|}$ for $i\ne j$ and $\sigma = 3$. We also consider a more difficult covariance structure, where $\rho_{i,j} = 0.9^{|i-j|}$. The detailed results are discussed in the Online Supplement.

\bigskip
\noindent
{\bf Example 3}. We demonstrate that our method performs well when the Gaussian assumption is not satisfied. To this end, we consider the same linear relationship as in Example~1, i.e.,
$Y =  3X_1-1.5X_2+2X_3+\sigma\,\veps$,
but the $X_j$s and $\veps$ are generated independently from the Student $t$ distribution with 5~degrees of freedom. We set $\sigma = 4$ and $\sigma = 8$ to make the signal to noise ratio comparable with Example~1.

\bigskip
The results for the three simulated examples are summarized in Tables~\ref{t:eg11}--\ref{t:eg3large}. In LARS-Corr, LASSO-Corr, permutation and truncated Gaussian tests-based methods, we take $\gamma \in \{0.01, 0.05, 0.2, 0.5\}$. Based on the simulation results, we can draw the following conclusions.

First, the test-based methods LARS-Corr and LASSO-Corr outperform the corresponding CV-based methods respectively for all scenarios, and the improvement of performance for our methods is more substantial when the signal is strong. Second, when the covariates are not equally correlated, our approach can still work well using~\eqref{eq:conv} as an approximation for the null distribution. Third, although LARS-Perm and LASSO-Perm have comparable performance to LARS-Corr and LASSO-Corr, respectively, they carry more computational costs. In addition, note that the permutation test can have much larger FP in some scenarios (e.g., LARS-Perm in Tables~\ref{t:eg12}--\ref{t:eg12large}). Fourth, although the truncated Gaussian tests have smaller false positives, their power is not very large. Therefore, the false negatives are still quite large even when $\gamma = 0.5$. As a result, the prediction errors are not well controlled. Finally, throughout the simulation experiments, the computational time of our methods drops dramatically compared with CV and permutation test.

\begin{figure*}[t!]
	\centering
	\subfigure[ $\sigma=6$ and $\rho=0$]{\includegraphics[width=.95\textwidth]{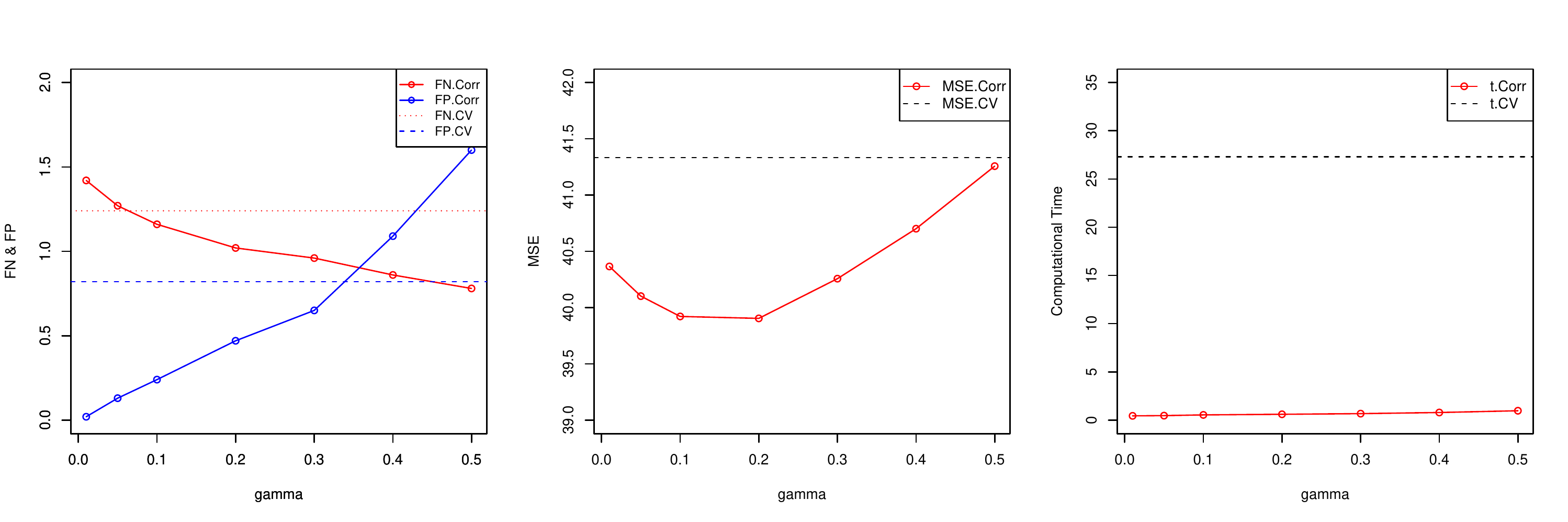}}
	\subfigure[$\sigma=6$ and $\rho=0.3$]{\includegraphics[width=.95\textwidth]{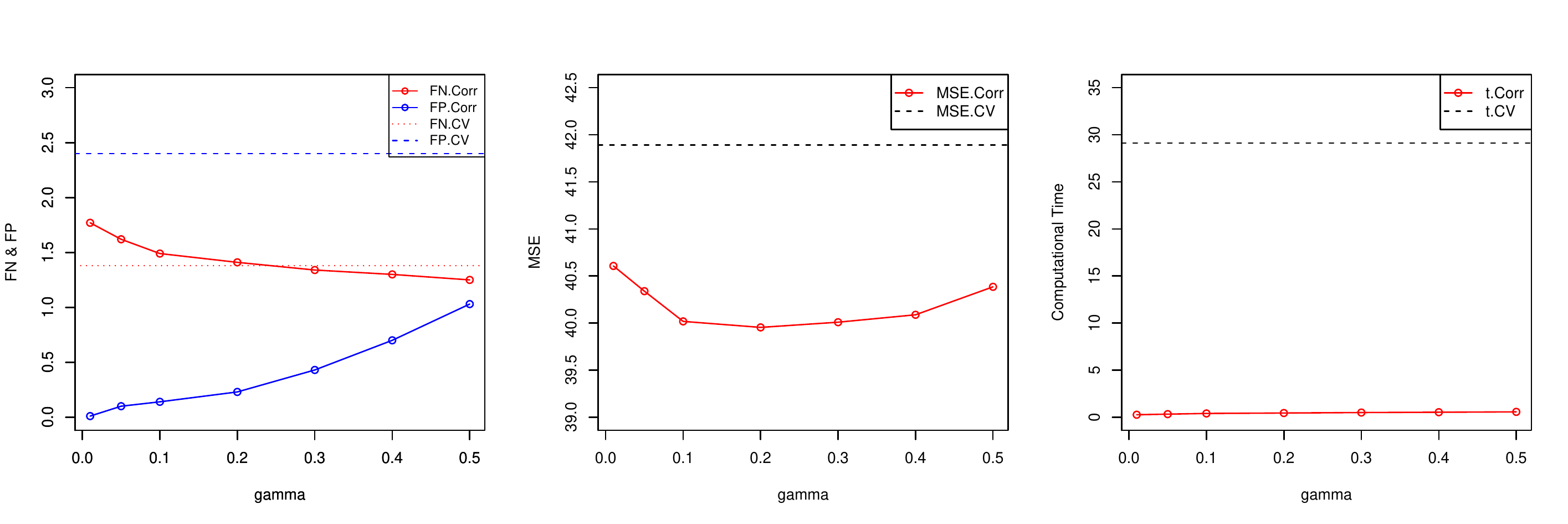}}
	\caption{Performance of LARS-Corr and LARS-CV in simulated example 1 with (a) $\sigma=6$ and $\rho=0$ and (b) $\sigma=6$ and $\rho=0.3$. In LARS-Corr, and $\gamma \in \{0.01, 0.05, 0.1, 0.2, 0.3, 0.4, 0.5\}$. For all three panels, the solid curve corresponds to LARS-Corr and the dashed curve corresponds to LARS-CV. In the first panel of (a) and (b), the red curves represent FN while the blue ones represent FP.}
	\label{fig:f1}
\end{figure*}
From Examples 1--3, one can see that our methods can control FN and FP by choosing a proper value of $\gamma$. We illustrate how the performance changes as the value of $\gamma$ varies for two scenarios in Figure~\ref{fig:f1}. This figure shows that as $\gamma$ increases, the FP of our methods has an increasing trend while the FN will decrease. Furthermore, our approach always outperforms CV in terms of MSE and computational time as $\gamma$ varies.

For independent cases, we also evaluate the performance of the proposed method using the maximal $t$-statistic described in \eqref{eq:tstat}. We find that the performance of our method with the maximal $t$-statistic is only slightly better than that with the maximal absolute correlation as the test statistic. Hence we do not include the detailed simulation results for the maximal $t$-statistic in this paper.

\subsection{A microarray data study}

We use a cardiomyopathy microarray dataset to demonstrate the performance of our method for high-dimensional problems. These data were previously analyzed in \cite{segal03, hall09, li12}. The aim of this study is to determine the most influential genes for a G protein-coupled receptor (Ro1) in mice. The dataset contains gene expression levels of 6320 genes on 30 specimens, in which the response variable is the expression level of Ro1 and the covariates $X_j$ are the expression levels of the remaining $p=6319$ genes.

\begin{table}[!t]
\scriptsize
	\centering
	\caption{The average MSE and computational time over 100 replications (with standard errors given in parentheses) for LARS-Corr, LARS-Perm, LARS-TG, LARS-CV, LASSO-Corr, LASSO-Perm, LASSO-CV, FSR-Corr, FSR-Perm, FSR-TG and FSR-CV on the gene expression data. For test-based approaches, $\gamma$ is set as 0.05, 0.1 and 0.2 respectively.}\label{t:real}
	{\vskip 2pt}	
	\begin{tabular}{lcrr|lcrr}
		Methods  & \multicolumn{1}{c}{$\gamma$}  & \multicolumn{1}{c}{MSE}  & \multicolumn{1}{c}{Time} & Methods  & \multicolumn{1}{c}{$\gamma$}  & \multicolumn{1}{c}{MSE}  & \multicolumn{1}{c}{Time}\\
		\hline
		\noalign{\vskip 2pt}
		LARS-CV   & & 0.63 (0.04) &  1.48 (0.02) & FSR-CV &  & 0.91 (0.16) & 0.78 (0.04) \\
		[2pt]
		LARS-Perm & 0.05 & 0.60 (0.05) & 1.81 (0.19) & FSR-Perm & 0.05 & 0.62 (0.05) & 1.44 (0.04) \\
		LARS-Perm & 0.1  & 0.59 (0.05) & 2.62 (0.35) & FSR-Perm & 0.1 & 0.63 (0.05) & 1.65 (0.06)\\
		LARS-Perm & 0.2 & 0.59 (0.04) & 5.78 (0.62) & FSR-Perm & 0.2 & 0.67 (0.05) & 2.22 (0.20) \\
		[2pt]
		LARS-Corr & 0.05 & 0.58 (0.05) & {\bf 0.44 (0.01)} & FSR-Corr & 0.05 & 0.61 (0.05) & {\bf 0.41 (0.01)}\\
		LARS-Corr & 0.1&  0.55 (0.05) & 0.53 (0.02) & FSR-Corr & 0.1 &  {\bf 0.60 (0.05)} & 0.48 (0.02) \\
		LARS-Corr & 0.2& {\bf 0.53 (0.04)} & 0.58 (0.03) & FSR-Corr & 0.2 &  {\bf 0.60 (0.05)} & 0.51 (0.02) \\
		[2pt]
		LARS-TG & 0.05 & 0.74 (0.05) & 3.42 (0.03) & FSR-TG & 0.05 & 0.72 (0.05) & 2.94 (0.02) \\
		LARS-TG & 0.1 & 0.72 (0.05) & 3.52 (0.03) & FSR-TG & 0.1 & 0.71 (0.05) & 3.01 (0.02) \\
		LARS-TG & 0.2 & 0.66 (0.05) & 3.56 (0.03) & FSR-TG & 0.2 & 0.65 (0.05) & 3.04 (0.02) \\
		[6pt]
		LASSO-CV   &  & 0.59 (0.04) & 1.98 (0.02) \\
		[2pt]
		LASSO-Perm & 0.05 & 0.60 (0.05) & 1.47 (0.05) \\
		LASSO-Perm & 0.1 &  0.57 (0.05) & 3.21 (0.60) \\
		LASSO-Perm & 0.2 &  0.54 (0.04) & 7.84 (0.99) \\		
		[2pt]
		LASSO-Corr & 0.05 & 0.58 (0.05) & {\bf 0.41 (0.01)} \\
		LASSO-Corr & 0.1  & 0.55 (0.05) & 0.49 (0.02)  \\
		LASSO-Corr & 0.2 & {\bf 0.53 (0.04)} & 0.55 (0.03) \\		
		\hline
	\end{tabular}
\end{table}

As in simulation studies, we perform all the methods, i.e., LARS-Corr, LASSO-Corr, FSR-Corr, LARS-Perm, LASSO-Perm, FSR-Perm, LARS-TG, FSR-TG, LARS-CV, LASSO-CV and FSR-CV on the dataset. For CV-based methods, we use 5-fold CV to implement model selection. As the average of pairwise correlations among covariates is close to 0 (less than 0.003), we use the null distribution for independent covariates in our test-based approaches. Since the correlation structure of the covariates in the gene expression data is different from iid Gaussian random variables, we also implement the permutation tests incorporated into LARS, LASSO and FSR correspondingly. In addition, we consider $\gamma \in \{0.05, 0.1, 0.2\}$ for LARS-Corr, LASSO-Corr, FSR-Corr, LARS-Perm, LASSO-Perm, FSR-Perm, LARS-TG and FSR-TG. In the experiment, 100 replications are conducted. For each replication, we randomly select 20 samples as the training data, and the remaining 10 as test data to obtain out-of-sample MSE.

We report the average of MSE and computational time with standard errors in Table~\ref{t:real}. One can see that our test-based methods using theoretical distribution have better prediction accuracy than CV-based ones. While permutation test has competitive performance for MSE, it has the most expensive computational cost among all methods. On the contrary, compared with CV as well as permutation test, the computational expenses of our test-based approaches are reduced for all three sequential selection procedures.

\begin{figure*}[t!]
	\centering
	\subfigure[Stepwise $p$-value and MSE]
	{\includegraphics[width = 0.48\textwidth]{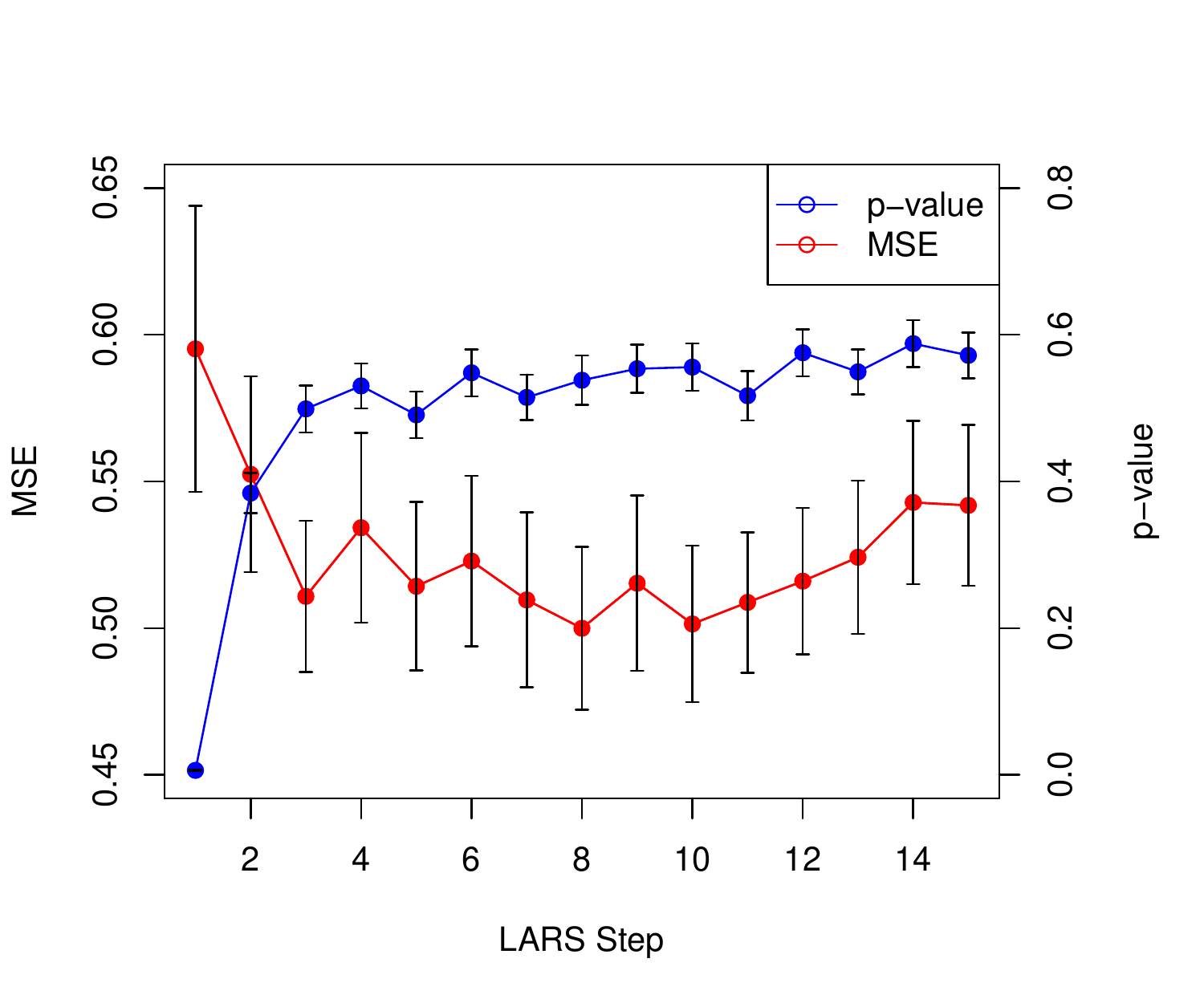}\label{fig:2a}}
	\subfigure[Most frequently identified genes and MSE]
	{\includegraphics[width = 0.48\textwidth]{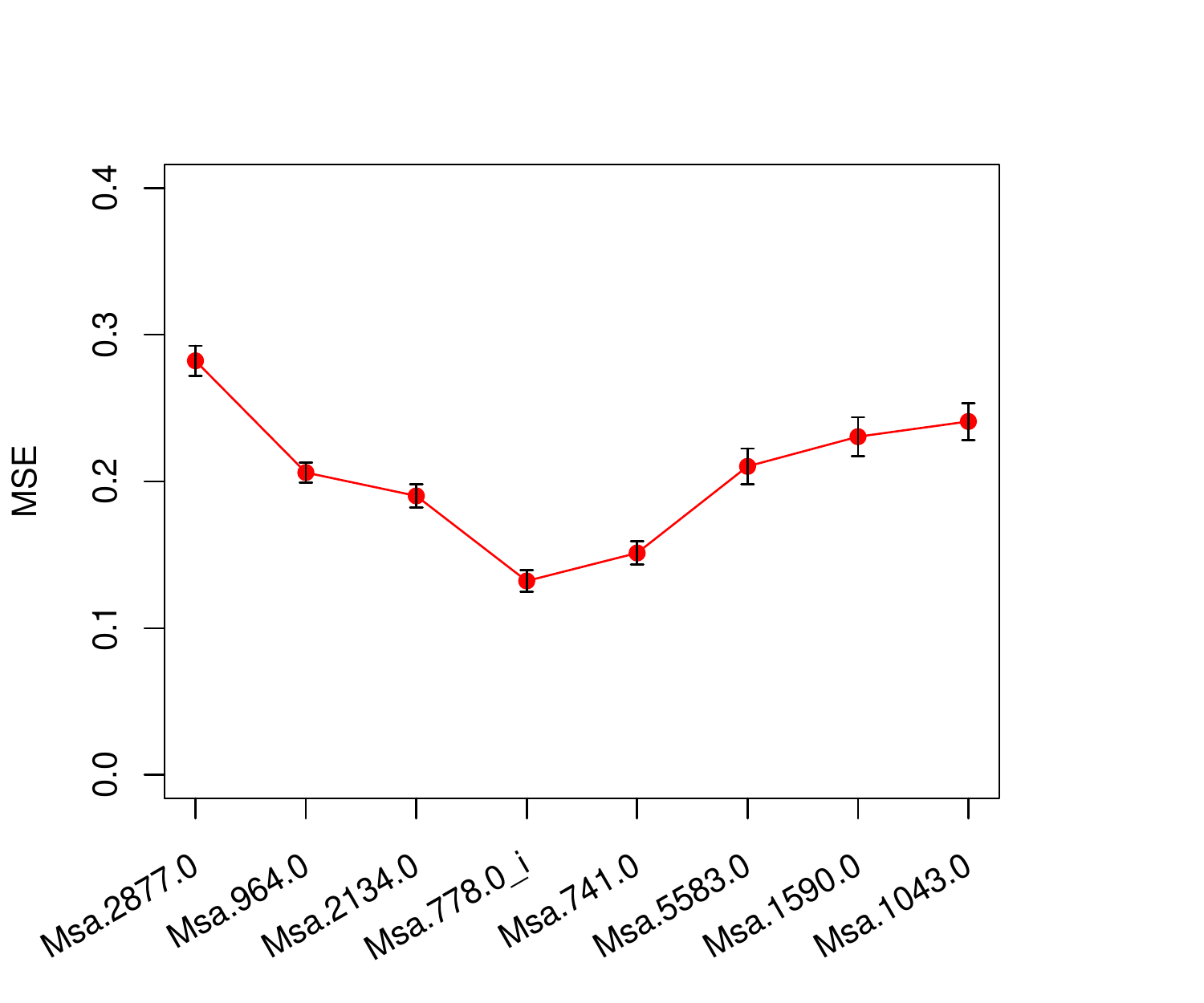}\label{fig:2b}}
	\caption{Performance of LARS-Corr applied to the microarray data. (a) Average $p$-value and MSE with one standard error bands for the first 15 steps of LARS-Corr over 100 replications. (b) 8 most frequently identified genes by LARS-Corr and the out-of-sample MSE corresponding to the model consists of the first $k \in \{ 1, \ldots, 8\}$ genes.}	\label{fig:real2}
\end{figure*}

To better demonstrate the performance of our test-based approach, we show a stepwise plot and an overall MSE plot for LARS-Corr as in Figure \ref{fig:real2}. Figure \ref{fig:2a} illustrates the stepwise $p$-value and MSE for the first 15 steps of LARS-Corr. Here the out-of-sample MSE at Step $k$ is with respect to the model containing variables selected by the first $k$ LARS steps. Note that such models might vary through 100 replications, resulting in relatively large standard errors for MSE. By the one standard error rule, Figure \ref{fig:2a} implies that a candidate model of size $3$ would be preferable. Moreover, we also summarize the most frequently identified genes out of 100 replications and sort by frequency from high to low. Figure \ref{fig:2b} shows the eight most frequently identified genes that are selected at least 10 times over 100 replications, as well as the out-of-sample MSE corresponding to the model containing the first $k$ genes with $k \in \{ 1, \ldots, 8\}$. Among the eight genes, Msa.2877.0 was also identified in \cite{hall09, li12}, and Msa.2134.0 was discovered in \cite{li12}. Overall, our variable selection method is effective in identifying potential scientific discoveries.

\section{Discussion}\label{sec:dis}

In this paper, we propose a test-based variable selection approach in the context of high-dimensional linear regression model with Gaussian covariates. We first formulate the null hypothesis, where we assume that the response is uncorrelated with all of the remaining covariates given a set of selected variables. We also propose the maximal absolute sample partial correlation statistic and discuss its asymptotic null distribution and power. We then incorporate the distribution information with sequential selection procedures. We use three simulated examples and one real data analysis to demonstrate that compared with CV-based procedure, the proposed method can perform variable selection effectively and efficiently.

Our proposed method involves sequential hypothesis testing. Therefore, instead of using a constant test level $\gamma$, one can consider multiple testing methods, such as the false discovery rate~(FDR) control \citep{bh95}, which provides flexible test levels and meaningful probability statements of the selected model. However, due to the adaptive nature of the sequential selection procedures, classical FDR control methods cannot be applied directly. There are some recent papers for sequential testing \citep{fs08alpha, ar14, gsell15seq}. However, the approaches in \cite{ar14, fs08alpha} are known to control the marginal FDR instead of the FDR. In contrast, \cite{gsell15seq} assumes that the $p$-values corresponding to the null hypotheses are iid $\mathcal{U}(0,1)$, which does not usually hold in our setting. We plan to investigate our procedure along this direction in future work.

\section*{Acknowledgments}
The authors were supported in part by NIH R01GM126550 and P01 CA-142538, and NSF grants IIS-1632951, DMS-1613112 and IIS-1633212. The authors thank the editor Dr. Christian Genest, the associate editor and reviewers for very helpful suggestions which led to substantial improvements of the paper.

\end{document}